\documentclass[twocolumn,showpacs,prl]{revtex4}
\usepackage{amsmath}
\usepackage{epsf}
\usepackage{graphicx}
\usepackage{dcolumn}
\usepackage{bm}

\begin{document}

\title{Nonanalytic Magnetic Response of Fermi- and non-Fermi Liquids}
\author{Dmitrii L. Maslov$^{1},$ Andrey V. Chubukov$^{2},$ and Ronojoy Saha$^{1}$}
\date{\today}

\begin{abstract}

We study the nonanalytic
behavior of the static spin
susceptibility of 2D fermions as a function of temperature and magnetic field.
For a generic Fermi liquid, $\chi_s (T, H)= {\rm
const}+c_1\max\{T,\mu_B|H|\}$, 
where $c_1$ 
 is shown to be expressed via
complicated combinations of the Landau parameters, rather than
 via the backscattering amplitude, contrary to the case of the specific
heat. Near a ferromagnetic quantum critical point, the field
dependence acquires a universal form $\chi^{-1}_s(H)={\rm
const}- c_2|H|^{3/2}$, with $c_2>0$. This behavior implies a
first-order transition into a ferromagnetic state.
 We establish a criterion for such a transition to win over the
transition into an incommensurate phase.
\end{abstract}

\affiliation{ $^{1}$Department of Physics, University of Florida,
P. O. Box 118440, Gainesville, FL
32611-8440 \\
$^{2}$ Department of Physics, University of Wisconsin-Madison,
1150 University Ave., Madison, WI 53706-1390} %\tableofcontents

\pacs{71.10. Ay, 71.10 Pm}
\maketitle

The nonanalytic behavior of thermodynamic quantities of a Fermi Liquid (FL)
has attracted a substantial interest over the last few years. The Landau
Fermi-liquid theory states that the specific heat coefficient $\gamma
(T)=C(T)/T$ and uniform spin susceptibility $\chi _{s}(T,H)$ of an
interacting fermionic system approach finite values at $T,H=0$, as in a
Fermi gas. However, the temperature and magnetic field dependences of $%
\gamma (T,H)$ and $\chi _{s}(T,H)$ turn out to be nonanalytic. In two
dimensions (2D), both $\gamma $ and $\chi _{s}$ are linear rather then
quadratic in $T$ and $\left| H\right|$ \cite{all}. In addition, the
nonuniform spin susceptibility, $\chi _{s}\left( q\right) ,$ depends on the
momentum as $\left| q\right| $ for $q\rightarrow 0$ \cite{bkv97,chm}.

Nonanalytic terms in $\gamma $ and $\chi _{s}$
arise due to a long-range interaction between quasiparticles mediated by
virtual particle-hole pairs. A long-range interaction is present in a Fermi
liquid due to Landau damping at small momentum transfers and dynamic Kohn
anomaly at momentum transfers near $2k_{F}$ (the corresponding effective
interactions in 2D are $|\Omega |/r$ and $|\Omega |\cos (2k_{F}r)/r^{1/2},$
respectively). The range of this interaction is determined by the
characteristic size of the pair, $L_{\mathrm{ph}}$, which is large at small
energy scales. To second order in the bare interaction, the contribution to
the free energy density from the interaction of two quasiparticles via a
single particle-hole pair can be estimated in 2D as $\delta F\sim u^{2}T/L_{%
\text{ph}}^{2},$ where $u$ is the dimensionless coupling constant.
As $L_{\text{ph}}\sim v_{F}/T$ by the
uncertainty principle, $\delta F\propto
T^{3}$ and $\gamma \left( T\right) \propto T$. Likewise, at $T=0$ but in a
finite field a characteristic energy scale is the Zeeman splitting $\mu
_{B}|H|$ and $L_{\text{ph}}\sim v_{F}/\mu _{B}|H|.$ Hence $\delta F\propto
\left| H\right| ^{3}$ and $\chi _{s}\left( H\right) \propto |H|$.

 A
second-order calculation indeed shows~\cite{chm,cmgg,betouras} that $\gamma $
and $\chi _{s}$ do depend linearly on $T$ and $\left| H\right| $. Moreover,
the prefactors are expressed only via two Fourier components of the bare
interaction [$U(0)$ and $U(2k_{F})$] which, to this order, determine the
charge and spin components of the backscattering amplitude $\Gamma
_{c,s}\left( \theta =\pi \right) $, where $\theta $ is the angle between the
incoming momenta. Specifically,
\begin{equation*}
\frac{\delta \gamma (T,H)}{\gamma (0,0)} =-\frac{9\zeta (3)}{\pi ^{2}}\left[
\Gamma _{c}^{2}(\pi )+3\Gamma _{s}^{2}(\pi )f_{\gamma }\left( \frac{\mu
_{B}|H|}{T}\right) \right] \frac{T}{E_{F}}
\end{equation*}
\begin{eqnarray}
\frac{\delta \chi _{s}(T,H)}{\chi _{s}(0,0)} &=&2\Gamma _{s}^{2}(\pi )\frac{T%
}{E_{F}}f_{\chi }\left( \frac{\mu _{B}\left| H\right| }{T}\right)  \notag \\
\frac{\delta \chi _{s}(q)}{\chi _{s}(0)} &=&\frac{4}{3\pi }\Gamma
_{s}^{2}(\pi )\frac{|q|}{k_{F}},  \label{1}
\end{eqnarray}
where $\delta \gamma (T,H)=\gamma (T,H)-\gamma (0,0)$, $\delta \chi
_{s}(T,H)=\chi _{s}(T,H)-\chi _{s}(0,0)$, $\delta \chi _{s}(q)=\chi
_{s}(q)-\chi _{s}(0)$, $\Gamma _{c}(\pi )=(m/2\pi )\left[ U(0)-U(2k_{F})/2%
\right] $, $\Gamma _{s}(\pi )=-(m/4\pi )U(2k_{F}),$ $\gamma (0)=m\pi /3,$ $%
\chi _{s}(0)=\mu _{B}^{2}m/\pi ,E_{F}=mv_{F}^{2}/2$, $\mu _{B}$ is the Bohr
magneton, and the limiting forms of the scaling functions are $f_{\gamma
}(0)=f_{\chi }(0)=1$ and $f_{\gamma }(x\gg 1)=1/3$, $f_{\chi }(x\gg 1)=2x$.
(Regular renormalizations of the effective mass and $g-$ factor are absorbed
into $\gamma (0)$ and $\chi _{s}(0)$).  The second-order susceptibility
increases with $H$ and $q$, indicating a tendency either to a
metamagnetic/first-order ferromagnetic transition or to a transition into a
spiral state. These tendencies signal a possible breakdown of the
Hertz-Millis scenario of the ferromagnetic
QCP \cite{hertz_millis}.

Experimentally, a linear $T$ dependence of the specific heat coefficient has been
observed in thin films of $^{3}$He~\cite{saunders}. A linear increase of $%
\chi _{s}$ with magnetization (and thus $H)$ has been observed in a 2D GaAs
heterostructure \cite{stormer}. Since none of these experiments correspond
to the weak-coupling limit, there is obviously a need for a nonperturbative
treatment of nonanalytic terms.

It has recently been shown \cite{cmgg,chmm} that the second-order result for
$\gamma (T,0)$ in Eq.(\ref{1}) becomes exact once the weak-coupling
backscattering amplitudes $\Gamma _{c,s}(\pi )$ are replaced by the exact
ones. This implies that the $O(T)$ term in $\gamma $ is determined
exclusively by 1D scattering events embedded into the 2D space. In these
events, fermions with almost opposite momenta experience either a small or
almost $2k_{F}$ momentum transfer. It has been conjectured in Ref. \cite
{cmgg} that the nonanalytic part of the spin susceptibility can be
generalized in the same way, i.e., by replacing weak-coupling $\Gamma _{s}$
in Eq. (\ref{1}) by its exact value. The same result was obtained within the
supersymmetric theory of the Fermi liquid \cite{efetov},
 and in the analysis of the scattering amplitude
 in the Cooper chanel, $\Gamma^C_s (\theta)$~\cite{finn_new}.
 Since
an extension of the second-order result for $\delta \chi _{s}$ hints at
far-reaching consequences for the ferromagnetic QCP, it is important to
establish a general form of 
$\chi _{s} (T,H)$.

In this communication, we present a general analysis of the nonanalytic
behavior of the spin susceptibility in 2D. We show that, in contrast to the
specific heat case,
higher orders
in the interaction are \emph{not }% 
 absorbed into the renormalization of $\Gamma _{s}(\pi )$
(equal to $\Gamma^C_s (\pi)$),
 but give rise to
extra $O(T,H)$ terms, whose prefactors are given by infinite series of the
scattering amplitudes $\Gamma _{s}(\theta )$ and $\Gamma^C_s (\theta)$ 
at all angles.
That higher-order terms in $\Gamma^C_s (\theta)$ are relevant was noticed in  
Ref.\cite{finn_new}. We show that higher-order terms in $\Gamma_s (\theta)$ 
 are also present.  These terms are more important than higher powers of
$\Gamma^C_s (\theta)$ as the latter are logarithmically reduced at small $T$
 and $H$.
  When the
interaction is neither weak nor peaked in some particular channel, the total
prefactor of the $O(T,H)$ term may, generally speaking, be of either sign.
However, the universality is restored near a ferromagnetic QCP,
where the $
n=0$ partial component of $\Gamma _{s}(\theta ) $ diverges. We show that
near the QCP the inverse susceptibility $\chi _{s}^{-1}(T=0,H)$ behaves as $%
\xi ^{-2}-A|H|^{3/2}$, where $\xi $ is the correlation length. This
dependence is dual to the $\xi ^{-2}-B\left| q\right| ^{3/2}$ behavior of
the nonuniform susceptibility \cite{pepin_prl}. Signs of $A$ and $B$ are
 positive, so that the nonanalytic terms destroy a continuous
transition towards a uniform ferromagnetic state, and, depending on
parameters, the system undergoes either a second-order transition into a
spiral phase or a first-order transition into a ferromagnetic state.

The temperature and magnetic-field dependences of $\gamma (T,H)$ and $\chi
_{s}(T,H)$ are most straightforwardly obtained by evaluating the
thermodynamic potential at finite $T$ and $H$, $\Xi \left( T,H\right) ,$ and
then differentiating it twice with respect to $T$ or $H$, respectively. To
understand the difference between the spin susceptibility and the specific
heat, consider for a moment the case of a contact interaction: $U(q) \equiv U
$.  To second order in $U$, the thermodynamic potential $\Xi $ is expressed
via the convolutions of the polarization bubbles $\Pi \left( \Omega
_{m},q,H\right) $ (with opposite spin projections for $H\neq 0$). The
polarization bubble has a conventional form
\begin{eqnarray}
\Pi _{\uparrow \downarrow }\left( \Omega _{m},q\right) &=&-\frac{m}{2\pi }%
\left( 1-\frac{\left| \Omega _{m}\right| }{\sqrt{\left( \Omega _{m}-2i\mu
_{B}H\right) ^{2}+v_{F}^{2}q^{2}}}\right)  \notag \\
&=&-\frac{m}{2\pi }+\Pi _{\text{dyn}}.  \label{bubbleud2D}
\end{eqnarray}
For large momenta ($v_{F}q\gg |\Omega _{m}|\sim \mu _{B}|H|$), the dynamic
part $\Pi _{\mathrm{dyn}}$ behaves as $|\Omega _{m}|/q$. Consequently, the
momentum integral $\int d^{2}q\Pi _{\mathrm{dyn}}^{2}$ diverges
logarithmically and is cut at $q=\max \{|\Omega _{m}|,\mu _{B}|H|\}$.
Because of the logarithm, the subsequent summation over frequencies yields a
universal term $\Xi (T,H)\propto \max \{T^{3},(\mu _{B}|H|)^{3}\}$. More
precisely, $\Xi (T,H)\propto T^{3}p\left( \mu _{B}H/T\right) $, where $%
p(x\ll 1)=a+bx^{2}+\dots $ and $p(x\gg 1)\sim \left| x\right| ^{3}$.
Accordingly, $\delta \gamma (T)/\gamma (0)\sim \delta \chi _{s}(T)/\chi
_{s}(0)\propto T$, and $\delta \chi _{s}(H)\propto |H|$.

To second order, $\delta \gamma (T)$ and $\delta \chi _{s}(T,H)$ behave
similarly.
The difference between the two quantities shows up
 at higher orders in $U$.  
In the rest of the paper, we 
 consider only 
 scattering in the particle-hole channel. 
 As we have already mentioned, there are higher-order contributions from the
Cooper channel, but they are logarithmically small in  a generic
 Fermi liquid,
and nonsingular near a ferromagnetic instability. A  particle-hole 
contribution of order $n$ contains integrals of $%
\Pi ^{n}=\Pi _{\text{dyn}}^{n}+c_{n-1}\Pi _{\text{dyn}}^{n-1}+...,$ where $%
c_{n}$ are the constants. The nonanalytic part
% in the specific heat is
of $\gamma(T)$
is solely related to the logarithmic divergence of the momentum integral $\int
d^{2}q\Pi _{\mathrm{dyn}}^{2}\propto \log |\Omega _{m}|,$ because only the
log singularity ensures the nonanalytic result of the subsequent Matsubara
sum: $T\sum \Omega _{m}^{2}\log |\Omega _{m}|=$const-$O(T^{3})$. The
momentum integrals of $\Pi _{\text{dyn}}^{k}$ with $k>2$ are not
logarithmically divergent, and the subsequent frequency summation gives rise
only to regular, $T^{2},T^{4}$... powers of $T$ in $\gamma(T)$. As a result,
higher-order diagrams for $\gamma (T)$ only renormalize bare interaction
$U$ into a full backscattering amplitude. For $\chi _{s}(H)$, the situation
is different -- higher powers of $\Pi _{\text{dyn}}$ do contribute
additional $|H|^{3}$ terms to $\Xi$, and hence additional $|H|$ terms to the
susceptibility. Indeed, evaluating $\Xi (0,H)$ to third order in $U$ and
retaining only the contribution $\delta \Xi ^{(3,3)}\left( 0,H\right)$ from $%
\Pi _{\text{dyn}}^{3}$, we obtain
\begin{eqnarray}
&&\delta \Xi ^{(3,3)}\left( 0,H\right) = \frac{u^{3}}{6\pi ^{2}}\mathrm{Re}%
\int_{0}^{E_{F}}d\Omega _{m}\int_{0}^{\infty }dqq  \notag \\
&&\times\frac{\Omega _{m}^{3}}{\left[ \left( \Omega _{m}-2i\mu_B H\right)
^{2}+v_{F}^{2}q^{2}\right] ^{3/2}}=\frac{2}{3\pi }\frac{u^{3}}{v_{F}^{2}}\mu
_{B}^{3}\left| H\right| ^{3},  \label{ac1}
\end{eqnarray}
where $u=mU/(2\pi )$.  The momentum and frequency integrals in $%
\delta \Xi ^{(3,3)}$
come from the region $|\Omega _{m}|\sim v_{F}q\sim \mu _{B}|H|$. This implies
that the $U^{3}|H|^{3}$ term in (\ref{ac1}) appears by purely dimensional
reasons, and does not require the $q$-integral in $\Xi $ to be
logarithmically divergent.

We see that the $\left| H\right| $ terms in $\delta \chi _{s}(H)$ coming
from two and three or more dynamic bubbles correspond to physically distinct
processes. The distinction becomes important for a generic Fermi liquid. The
contribution to $\delta \chi _{s}(H)$ from two dynamic bubbles, which starts
at order $U^{2}$, is generalized beyond the weak-coupling limit by replacing
the bare interaction by a fully renormalized static vertex.
Using the same %
computational procedure as in Ref.\cite{chmm}, we find that,
similarly to the specific heat, the exact result for this
contribution is expressed in terms of the spin 
part  of the
backscattering amplitude $\Gamma _{s}\left( \pi \right) $:
\begin{equation}
\delta \chi_s^{(2)} (H)\rightarrow \delta \chi _{s}^{BS}=\frac{8}{\pi
v_{F}^{3}}\Gamma^2 _{s}\left( \pi \right) \mu _{B}^{3}\left| H\right| .
\label{ac2}
\end{equation}
The same result was obtained in Refs.~\cite{efetov,finn_new}. The
logarithmic divergence of the momentum integral of $\Pi _{\mathrm{dyn}}^{2}$
is the crucial element in the derivation of Eq.(\ref{ac2}), as the
higher-order corrections can be absorbed into
static $\Gamma_s (\pi)$ only
if typical $v_{F}q$ are much larger than
typical $|\Omega _{m}|$, given by $\mu _{B}|H|$.

On the other hand, contributions to $\delta \chi _{s}(H)$ from three and
more dynamic bubbles come from $v_{F}q\sim |\Omega _{m}|$,
 and are expressed via the convolutions of the partial harmonics of the
scattering amplitudes, which do not reduce to higher powers of the
backscattering amplitude. As an illustration, we consider a generalization
of the third-order contribution $\delta \Xi ^{(3,3)}$, assuming that the
spin component of the scattering amplitude has only two partial components: $%
n=0,1$, i.e., $\Gamma_{s}(\theta )=\Gamma _{s,0}+\cos \theta \Gamma _{s,1}$.
Replacing each interaction vertex by $\Gamma_{s}(\theta )$, we obtain
\begin{eqnarray}
&&\delta \chi ^{(3)}_{s}(H)\rightarrow\delta\chi^{\mathrm{any}}_{s}=-\frac{64%
}{\pi v_{F}^{3}} \mu _{B}^{3}\left| H\right|  \notag \\
&&\times\left[ \Gamma _{s,0}^{3} -a_1 \Gamma _{s,0}^{2}\Gamma _{s,1} - a_2
\Gamma _{s,0}\Gamma _{s,1}^{2}+ a_3 \Gamma _{s,1}^{3}\right] .  \label{ac5}
\end{eqnarray}
where $a_1 = 3\left( 2\ln 2-1\right),~a_2 = 3(3\ln 2-2),~ a_3 = \left( 5/2
-3\ln 2\right)$. This expression obviously does not reduce to the cube of
the backscattering amplitude, which in this approximation would be equal to $%
\Gamma _{s}(\pi )=\Gamma _{s,0}-\Gamma _{s,1}$. Higher-order contributions
are given by progressively more complicated combinations of $\Gamma_{s,0}$ and
$\Gamma_{s,1}$.

The total result for $\delta\chi_s$ is a sum of backscattering
 and all-angle scattering contributions. 
Since $\Gamma_s (\pi)$ is equivalent to the spin component of the 
particle-particle scattering amplitude 
$\Gamma^C_s (\pi)$ (Refs. \cite{chm,finn_new}), 
the repulsive interaction in the Cooper channel leads to a
logarithmic reduction of $\Gamma _{s}\left( \pi \right) $
\cite{aleiner_efetov,efetov,finn_new}.
  For $T=0,~H\rightarrow 0,$ $\Gamma _{s}\left(
\pi \right) \propto 1/\log \left| H\right| $ and therefore $\delta \chi
_{s}^{\mathrm{BS}}\propto \left| H\right| /\log ^{2}\left| H\right|$~\cite
{comm_3}. On the other hand, contributions to $\delta \chi _{s}$ from three
and more dynamic bubbles contain angular averages of $\Gamma _{s}\left(
\theta \right) ,$ which are not affected by the Cooper singularity.
Therefore, the $|H|$ terms from these contributions do not acquire
additional logarithms and win over the backscattering contribution  for $%
T,H\rightarrow 0$ 
(and also 
over higher-order terms in $\Gamma^C_s (\theta)$, which
 vanish logarithmically at $T, H \rightarrow 0$).
 Note that for $|\Gamma _{s,1}/\Gamma_{s,0}| <1$, the sign
of $\chi _{s}^{\mathrm{any}}$ %%
in Eq.~(\ref{ac5}) is determined just by the sign of $\Gamma _{s,0}$: for
negative $\Gamma _{s,0}$ (corresponding to enhanced ferromagnetic
fluctuations), $\delta \chi _{s}$ increases with $H$, whereas for positive
 $\Gamma _{s,0}$, $\delta \chi _{s}$ decreases with $H$.

Next, we discuss the behavior of the spin susceptibility in the vicinity of a
ferromagnetic QCP, where $\Gamma _{s,0}$ diverges, while other components of
$\Gamma _{s}$ remain finite. At any finite distance from the QCP, the
backscattering amplitude still vanishes as $1/\log \max \{\mu _{B}|H{|,}T{\}.%
}$ However, at large $\Gamma _{s,0}$, this behavior is confined to an exponentially
small range of $H$ and $T,$ which we will not consider below. Outside this
range, the backscattering amplitude diverges as $\Gamma _{s,0}$, and the
backscattering contribution $\delta \chi _{s}^{\mathrm{BS}}(T,H)$ diverges
as $\Gamma _{s,0}^{2}$. All-angle contributions, however, diverge even
stronger, and one needs to sum up a full series of diagrams to obtain the
behavior of $\delta \chi _{s}(T,H)$ near QCP. To do this, we assume, as it
was done in Refs.\cite{pepin_long,cmgg}, that the Eliashberg approximation
is valid near QCP because overdamped spin fluctuations are slow compared to
fermions.
In the Eliashberg theory, the field-dependent part of the thermodynamic
potential is $\Xi =(1/2\pi )T\sum_{\Omega _{m}}\int dqq\log \chi _{0}^{-1}(q,\Omega ,H)$,
where
\begin{equation}
\chi _{0}(q,\Omega _{m},H)=\frac{m}{\pi }\frac{\mu
_{B}^{2}}{\delta+a^{2}q^{2}+(2\pi/m)\Pi _{\text{dyn}}(\Omega
_{m},q)}, \label{ac7}
\end{equation}
is the dynamic spin susceptibility without nonanalytic corrections,
$\delta=|\Gamma_{s,0}|^{-1}$, and $a$ is the radius of the exchange
interaction, required to be large ($ak_F\gg 1$) for the Eliashberg
theory to work \cite{dzero}. $\Pi _{\mathrm{dyn}}$ differs from
Eq.(\ref{bubbleud2D}) in that (i) it is built on full Green's
functions (containing self-energies) and (ii) $\mu _{B}$ in the
denominator is replaced by $\mu _{B}^{\ast }=\mu _{B}/\delta$ (cf.
Ref. ~\cite{beal_monod}). To illustrate once again the difference
between the specific heat and spin susceptibility, we set $\xi
=\infty $ in the denominator of Eq.(\ref{ac7}) and neglect the
self-energy renormalization for a moment. Evaluating the
derivatives of $\Xi (T,H)$ with respect to $T$ and $H$, we find
that the prefactor of the $T$ term in the specific heat
coefficient diverges~\cite{chmm}, whereas the prefactor of the
$|H|$ term in the spin susceptibility remains finite: $\delta \chi
_{s}(H)/(\mu _{B}^{\ast })^{3}=(2/\pi v_{F}^{2})|H|$. 
This indicates dramatic cancellations between diverging terms in the
perturbation theory for $\delta \chi _{s}(H)$~\cite{comm_1}.

A complete result for the susceptibility near QCP is obtained by including
the self-energies when evaluating $\Pi _{\mathrm{dyn}}$ in Eq.(\ref{ac7})
(vertex corrections are small, see Ref. \cite{pepin_long}).  
The self-energy
near the QCP interpolates between $\Sigma =\lambda \omega _{m}$ away from
QCP and$\Sigma =\omega _{0}^{1/3}|\omega _{m}|^{2/3}$ near QCP, where
$\lambda =3/(4k_Fa\delta)$ and $\omega _{0}=3\sqrt{3}E_F/(4(k_Fa)^4) $
~\cite{aim,finn}.
 Using these expressions, we obtain for the inverse susceptibility
\begin{equation}
\chi _{s}^{-1}(H, T=0)\propto\delta-\frac{8}{3}\frac{\mu _{B}^{\ast }|H|}{%
v_{F}/a}~\sqrt{\delta} ~K_{H}\left(\frac{\mu _{B}^{\ast
}|H|ma^2}{\delta}\right) \label{ac9}
\end{equation}
where $K_{H}(0)=1$ and  $K_{H}(x\gg 1)=1.25 \sqrt{x}$, The limit
of $x\rightarrow \infty $ describes the situation right at QCP.
Here, divergent $\xi$ cancels out from the answer, and the $H$
dependence of $\chi_s ^{-1}$ becomes $|H|^{3/2}.$ We emphasize
that the exponent of $3/2$ is the consequence of the non-Fermi
liquid behavior, manifested by the divergence of the ``effective
mass'' $\partial \Sigma /\partial \omega _{m}\propto \omega
_{m}^{-1/3}.$

The nonanalytic $\left| H\right| ^{3/2}$ dependence exists only at $%
T\rightarrow 0$. At finite $T$, the field dependence of the spin
susceptibility is analytic: $\delta \chi _{s}\propto H^{2}$. However, the
prefactor scales as $1/(\lambda T)$ away from QCP, and as $T^{-1/6}$ at QCP.
AT $H=0$, $\delta\chi_s(T)\propto T\log T$ \cite{pepin_long,comm_2}.

A complementary way to see the nonanalytic dependence the susceptibility on
the magnetic field is to analyze the thermodynamic potential itself. Viewed
as a function of the magnetization %%
$M=(m/\pi) \eta \mu _{B} $, where $2\eta $ is the difference of the Fermi
energies for spin-up and spin-down fermions, the thermodynamic potential $%
\Xi (T=0,\eta )$ contains a nonanalytic $|\eta |^{3} $ term away from
criticality: $\Xi _{na}(0,\eta )=-|\eta |^{3}/48\pi v_{F}^{2}\lambda .$ Near
QCP, $\lambda $ diverges and the $|\eta |^{3}$ dependence is replaced by $%
|\eta |^{7/2}$, in agreement with the $H^{3/2}$ field dependence of the
spin susceptibility. At finite $T$, the $|\eta|^{7/2}$ term
evolves into an analytic $\eta^4$ one with a singular prefactor
$T^{-1/6}$.

We now study the consequences of the nonanalytic behavior of $\chi
_{s}(T,H,q)$. First, we see from Eq.(\ref{ac9}) that the spin
susceptibility diverges at some finite value of $H$, which implies
that a second-order ferromagnetic QCP is preempted by the
first-order one. This possibility
 was discussed in detail in Ref.\cite{bk} -- our analysis differs
 from this work
in that we include the fermionic self-energy and nonanalytic $T$ dependence of
the susceptibility. Assuming that the first-order transition occurs near
QCP, where the nonanalytic term in $\Xi (T=0,\eta )$ is $\eta ^{7/2}$, we
have
\begin{equation}
\frac{\pi }{m}~\Xi (T=0,\eta )=\frac{\delta}{2}\eta^2-\frac{|\eta
|^{7/2}}{E^{3/2}}+b^2\eta ^{4},  \label{b}
\end{equation}
with
$E=3.82E_F/(k_Fa)^{4/3}$.
 Because of the nonanalytic term, $\Xi $ has a minimum at
finite $\eta $. The
first-order transition occurs at
$\delta_H=\left(k_Fa/3.32\right)^8/(bE_F)^6$ when $\Xi =0$ at
this minimum.  By order of magnitude, $b\sim 1/E_{F}$.
The first-order transition occurs in the critical
region ($\delta_H<1$) for $k_Fa<3.32 
(b E_F)^{3/4}$.

The susceptibility at $H=0$ but finite $q$ is given by
\begin{equation}
\chi _{s}^{-1}(q)= \delta+a^2\left(
q^{2}-cq^{3/2}k_{F}^{1/2}\right), \label{dm1}
\end{equation}
where $c\approx 0.25$ \cite{pepin_long}. $\chi _{s}^{-1}(q)$
diverges at $q=q_0=0.035k_F$
for $\delta_q=0.42\times 10^{-3}(ak_F)^2$. This signals a transition
into an incommensurate phase.

\begin{figure}[tbp]
\begin{center}
\epsfxsize=0.5\columnwidth
\epsffile{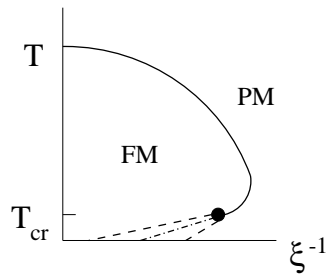}
\end{center}
\par
\caption{Schematic phase diagram near a ferromagnetic QCP for
$\protect\rho<1 $ (see text).
$\protect\delta$ is a control
parameter, e.g., pressure.
The second-order phase transition
becomes first-order below a tricritical point because of the
nonanalytic $\protect\eta^{7/2}$ term in Eq.~(\ref{b}). At small
$T$, the transition line has an $S$-shaped form because of the
negative $T$ dependence of $\chi_s^{-1}(T)$.} \label{fig1}
\end{figure}

Which of the two instabilities occurs first depends
 on the nonuniversal parameter
 $\rho= \delta_q/\delta_H  = (1.35 b E_F/ak_F)^6$.
 For $\rho >1$, the first instability is into the incommensurate phase; for $%
\rho <1$, the first-order transition occurs first (see Fig.
\ref{fig1}). Although formally 
$a k_F$ should be large, 
 both situations are actually possible, 
particularly if $b E_F$ is a large number.

At finite $T$, the transition line has an $S-$shaped form (see Fig. \ref{fig1})  because
of the negative $T$ dependence of $\chi ^{-1}(T)$ 
\cite{comm_4}.
The tricritical point 
%at $T=T_{cr}$,
 separating the second- and first-order
transitions,  results from the balance
 between the $b^2 \eta^4$ term in Eq.~(\ref{b}) and the  
$\eta^4/T^{1/6}$ term 
which replaces the $|\eta|^{7/2}$ term at a finite $T$.

To summarize, in this paper we considered the temperature and magnetic field
behavior of the spin susceptibility of a 2D Fermi liquid, both away and near
a ferromagnetic QCP. We found that in a Fermi-liquid phase, $\delta \chi
_{s}(T,H)\propto \max \{T,|H||]$, but the prefactor is not expressed solely
in terms of the backscattering amplitude, in contrast to the specific heat.
At a ferromagnetic QCP, the magnetic field dependence of $\chi ^{-1}(T=0,H)$
becomes $H^{3/2}$, with the universal, negative prefactor. This behavior
favors a first-order transition to ferromagnetism and competes with $q^{3/2}$
behavior of $\chi ^{-1}(T=H=0,q)$ which favors an incommensurate spin
ordering.

We acknowledge helpful discussions with I. Aleiner, K. Efetov, A.
Finkelstein, G. Schwiete, A. Millis and T. Vojta,
support from NSF-DMR 0240238 (A. V. Ch.), NSF-DMR-0308377 (D. L. M.),
and the hospitality of the Aspen Center of Physics.


\begin{thebibliography}{99}
\bibitem{comm_3}
 The formula in the text, $\Gamma_{s} (\pi) =\Gamma^C_s (\pi) 
\propto 1/\log H$, is
valid when the sum of partial components for $\Gamma^c_s$  converges, 
otherwise $\Gamma_{s} (\pi)$ has extra dependence on $H$, e.g.,
$\Gamma_{s} (\pi) \propto \log {\log H}/\log H$ for 
 a model 
used in ~\cite{aleiner_efetov}.
\bibitem{comm_1}  Away from the QCP, we obtain in this approximation
 $\delta \chi _{s}(H)=(2/(\pi v_{F}^{2}))(\mu _{B}^{\ast
})^{3}|H|F_{H}\left(\delta^{-1} \right) $, where $%
F(x)=(x/(1+x))^{2} $. The small $x$ limit agrees with the perturbation
theory; at large $x$, $F(x)$ tends to a constant.
\bibitem{comm_2}  Away from QCP, we obtain,  neglecting the fermionic
self-energy,
$\delta\chi^{-1}_s (T, H=0)
%\propto \xi^{-2}
\propto
 - (T/E_F)~ F_T
\left(\delta^{-1} \right)$, where $F_T (x) = \log(1+x) - x/(1+x)$
At small $x$, $F_T (x) \approx x^2/2$, in agreement with the
perturbation theory. At large $x$, $F_T (x) \approx \log x $.
\bibitem{comm_4}
We assume that the real
system is quasi two-dimensional, 
 so that there is a real phase transition 
 at a finite $T$.
\end{thebibliography}
\end{document}